\documentclass{article} 
\usepackage{graphicx} 
\usepackage{multirow}
\pdfoutput=1
\usepackage{multicol}
\usepackage{float}
\usepackage{amsmath}
\usepackage{enumitem}
\usepackage{amssymb}
\usepackage{amsthm}
\usepackage{pifont}
\usepackage{csquotes} 
\usepackage{booktabs}
\usepackage[margin=1in]{geometry}
\usepackage[english]{babel}
\usepackage{bigstrut}
\usepackage{colortbl}
\usepackage[skip=5pt, indent=20pt]{parskip}
\usepackage{subcaption, tablefootnote}
\usepackage{comment}
\usepackage[
backend=biber,
sorting = nty, 
style=apa
]{biblatex}

\theoremstyle{definition}

\newpage
\begin{document}
\begin{titlepage}
    \begin{center}
        \vspace*{1cm}
        
        \Huge
        \textbf{Evaluating Fairness of Voting Systems: Simulating Violations of Arrow’s Conditions}
        
        \vspace{1.5cm}
        \LARGE
        \textbf{Virochan Pandit}    vpandit3@jhu.edu\\
        Johns Hopkins University\\
        \vspace{0.5cm}
        \textbf{Joseph Cutrone}     jcutron2@jhu.edu\\
        Johns Hopkins University 
        \vfill
    
        \LARGE
        This research was supported by Johns Hopkins'\\
        Provost's Undergraduate Research Award (PURA)\\

        \vspace{1cm}
        
        \normalsize
        We would like to acknowledge Christopher Ratigan and Sergey Kushnarev for their insights and suggestions on this paper. We also thank Alexander Shumakovitch for guidance on operating the Rockfish computer cluster. Further, we thank the Hopkins Office for Undergraduate Research (HOUR) for their support and guidance.

    \end{center}
\end{titlepage}

\frenchspacing

\pagebreak
\begin{abstract} 
This paper builds upon the work of Dougherty and Heckelman (2020) by determing the frequency that 13 voting systems violate Arrow's social choice criteria with up to six alternatives. These results determine which of the 13 voting systems, is the fairest based on their probabilistic likelihood of violating Arrow's  social choice criteria. The voting systems considered are: Plurality, Borda, Dowdall, Top Two, Hare, Coombs, Baldwin, Copeland, Anti-Plurality, Nanson, Ranked Pairs, Pairwise Majority, and Minimax. Elections with up to 10,000 voters and between three and six alternatives are simulated using both Impartial Culture and Impartial Anonymous Culture. These simulations show that Pairwise Majority is the least likely to jointly violate Arrow's criteria. As the number of alternatives increases, the joint-violation frequencies increase for each voting method. For all systems except Pairwise Majority, the joint-violation frequencies in elections with at least 30 voters and four alternatives are greater than 98\%.

\end{abstract}
\pagebreak


\pagebreak
\section{Introduction}
Social choice theory studies how individual preferences can be aggregated into collective decisions, often through voting systems. A central challenge in this field is determining whether any such system can satisfy desirable fairness criteria. Kenneth Arrow’s Impossibility Theorem (1963) famously demonstrated that no rank-order voting system can satisfy a set of intuitive fairness conditions - namely, Non-Dictatorship, Unrestricted Domain, Transitivity, Unanimity, and Independence of Irrelevant Alternatives (IIA) - when there are at least three alternatives and two or more voters. While the theorem is negative in nature, it has motivated decades of research focused on evaluating which voting systems are least likely to violate these, and other, criteria in practice, especially under probabilistic models of voter behavior (e.g., \cite{Gehrlein1994-GEHTEL}; \cite{Arrow}).

Building on this literature, this paper uses large-scale simulations to estimate the frequency with which commonly studied voting systems violate Arrow’s criteria. Our work extends the recent empirical analysis of Dougherty and Heckelman (2020), who examined the probability of criterion violations for elections with three alternatives. We extend their results to elections with up to six alternatives, using both the Impartial Culture (IC) and Impartial Anonymous Culture (IAC) probability distributions to generate preference profiles. 

Our analysis quantifies the probability that each system satisfies Arrow’s criteria individually and jointly. We introduce a joint-violation metric as a benchmark of fairness: the likelihood that a system violates Arrow's criteria in a given election. This measure enables meaningful comparison across systems.

\noindent The main findings are:
\begin{itemize}
\item Of all systems, Pairwise Majority is the least likely to jointly violate Arrow’s criteria.

\item Among systems that guarantee transitive outcomes, the Ranked Pairs method has the lowest joint-violation frequencies. However, under most settings these frequencies are greater than 70\%.

\item As the number of alternatives increases, joint-violation frequencies increases sharply for any number of voters.
\end{itemize}

This study contributes to the ongoing normative evaluation of voting systems by highlighting how violation frequencies behave in large-scale simulations and across a wider range of alternatives. It provides evidence that while no voting system is perfect, some perform significantly better than others in satisfying foundational fairness criteria. Furthermore, we improve on existing methodology by using a more sensitive algorithm for detecting IIA violations, which yields greater lower bounds and helps reconcile discrepancies in prior results.

The remainder of the paper is structured as follows: Section 2 defines the voting systems under consideration. Section 3 reviews Arrow’s fairness criteria. Section 4 explains the simulation methods. Section 5 presents the main results and Section 6 summarizes our conclusions. Section 7 contains the author's declaration statements. 

\section{Voting Systems}
Let $C = \{ 1, 2, \dots, m \}$ be a finite set of alternatives and $V = \{ 1, 2, \dots , v \}$ be a finite set of voters. 

We consider the same set of voting systems as defined by Dougherty and Heckelman (2020). Following their lead we classify Plurality, Minimax, Top Two, Hare, Anti-Plurality, and Nanson as \textbf{Group 1} because those voting systems always violate IIA and can violate Pareto - two of Arrow's conditions defined below. 

We adopt the following notation: for two alternatives \(A\) and \(B\), \(A \succ B\) indicates society prefers \(A\) over \(B\), \(A \succeq B\) indicates society prefers \(A\) at least as much as \(B\), and \(A \sim B\) indicates society is indifferent between \(A\) and \(B\).

\begin{description}

    \item [Plurality -]
    Alternatives are ranked in descending order based on their first-place vote tally. Alternatives with the same number of first place votes are given the same rank. 

    \item [Anti-Plurality -]
    Alternatives receive zero points for every last place vote and 1 point for all other place votes. Alternatives are ranked in descending order based on their point total. Alternatives 

    \item [Borda Count -] 
    For each 1st place vote, alternatives receive \(m\) points, for each 2nd place vote they receive \(m - 1\) points, and so on until for an \(m\)th place vote, alternatives receive one point. Alternatives are ranked in descending order based on their point total. 

    \item [Dowdall -]
    For each first place vote, alternatives receive 1 point, for a second place vote they receive \(\frac{1}{2}\) point, for a third place vote they receive \(\frac{1}{3}\) point and so on until for an \(m\)th place vote, alternatives receive \(\frac{1}{m}\) point. Alternatives are ranked in descending order based on their point total. 
    
    \item [Top Two/Plurality Runoff -]
    Begin with plurality. Eliminate all alternatives except the two with the most plurality points. Take a second plurality vote on just those two alternatives and then eliminate the alternative with fewer plurality points. Rank alternatives in reverse order of elimination. (ensuring the alternative(s) eliminated last is most preferred in the social ranking). 
    The remaining alternatives are ranked in descending order. 

    \item [Hare -]
    Begin with plurality. Eliminate the alternative with the fewest plurality points (eliminating all alternatives that tie for fewest points). Repeat the process on the remaining alternatives until all alternatives are eliminated. Rank alternatives in reverse order of elimination (ensuring the alternative(s) eliminated last is most preferred in the social ranking). 
 
    \item [Coombs -]
    Eliminate the alternative(s) ranked last by the most voters (eliminating all alternatives that tie for most last place votes). Repeat the process on the remaining alternatives until all alternatives are eliminated. Rank alternatives in reverse order of elimination (ensuring the alternative(s) eliminated last is most preferred in the social ranking). 

    \item [Baldwin -]
    Begin with Borda. Eliminate the alternative with the fewest Borda points (eliminating all alternatives that tie for fewest points). Repeat the process on the remaining alternatives until all alternatives are eliminated. Rank alternatives in reverse order of elimination (ensuring the alternative(s) eliminated last is most preferred in the social ranking).

    \item [Nanson -]
    Begin with Borda. Eliminate any alternative(s) at or below the mean of Borda points. Repeat the process on the remaining alternatives until all alternatives are eliminated. Rank alternatives in reverse order of elimination. (ensuring the alternative(s) eliminated last is more preferred in the social ranking).
    
    \item [Pairwise Majority -]
    Let $A,B \in C$ be two distinct alternatives. The head-to-head comparison between $A$ and $B$ is determined by how many voters rank $A$ above $B$ compared to how many rank $B$ above $A$. If $A$ defeats $B$ when compared head-to-head, then \(A \succ B\). If \(B\) defeats $A$ when compared head-to-head then \(B \succ A\). If  they tie when compared head-to-head, then \(A \sim B\). Rank alternatives based on the pairwise comparisons.
    
    \item [Copeland/Pairwise Comparison -]
    All alternatives are compared head-to-head pairwise. For each comparison, the alternative that wins head-to-head receives 1 point. If the two alternatives tie they receive \(\frac{1}{2}\) point each. Rank alternatives in descending order of total points.

    \item [Minimax/Simpson-Kramer -]
    All alternatives are compared head to head. For each alternative, their score is the magnitude of their worst pairwise defeat. Rank alternatives in reverse order of the magnitude of the worst defeats.

    \item [Ranked Pairs -]
    Begin with pairwise majority and sort pairwise preferences in descending order of vote margins. Lock in each pair starting with the pair that has the greatest vote margin as long as it does not create a cycle.
\end{description}

\subsection{Breaking Ties}
In Plurality, Anti-Plurality, Borda Count, Dowdall, Copeland, and Minimax, alternatives with the same number of points or votes are assigned the same rank. In round based systems, such as Hare, Coombs, Baldwin, and Nanson, alternatives eliminated in the same round are assigned the same rank. In Pairwise Majority, if $A$ and $B$ are tied in the head-to-head comparison then $A \sim B$ in the final societal ordering. For Ranked Pairs, if two or more pairs share the largest vote margin, one pair is randomly selected. For Top Two, a tie may occur in the first round if more than two alternatives receive an equal number of votes. In this case, ties are broken randomly. 
  
\subsection{Arrow's Criteria of Fairness}
Following \parencite{Arrow, sep-arrows-theorem,pennpaper}, below is a list of Arrow's voting criteria used in this paper. 

\begin{enumerate}

    \item \textbf{Unrestricted Domain (UD)} -
    The domain of the voting system is every possible preference schedule.

    \item \textbf{Non-Dictatorship (ND)} - 
    There is no individual whose preferences uniquely determine the social raking of all alternatives, regardless of how other individuals rank the alternatives. 

    \item \textbf{Transitivity (TR)} - $\forall a,b,c \in C, a \succeq b, b \succeq c \implies a \succeq c$.

    \item \textbf{(Pareto) Unanimity (U)} - If every individual prefers $a$ to $b$, then $a \succ b$ in the final societal ranking.

    \item \textbf{Independence of Irrelevant Alternatives (IIA)} - Society's preference between any two alternatives only depends on voter's preferences between these two alternatives \parencite{HK,MathCB}. Equivalently, if there are two preference schedules in which each voter has the same relative preference between alternatives \(A\) and \(B\), the societal preference between \(A\) and \(B\) must be the same in both preference schedules. 
\end{enumerate}

\section{Arrow's Theorem}
\newcommand{\cmark}{\checkmark}
\newcommand{\xmark}{\ding{53}}
Arrow’s theorem shows that no voting system can satisfy all five conditions in a society with at least two voters and at least three alternatives  \parencite{mainArrow}. Table 1 below summarizes which criteria each voting system satisfies assuming an unrestricted domain (UD). A check $\cmark$ indicates that the voting system always satisfies the criteria while an $\times$ indicates that the system may violate that criterion under an UD. The theoretical properties of voting systems are well studied in literature \parencite{HK, JSTORp1, Bordaclc,Nansonnote}. 

\begin{table}[H]
\caption{Criteria satisfied by 13 Voting Systems}
\centering
\begin{tabular}{ c | c c c c c c c c c c} 
 Voting System & ND & TR & U & IIA \\
 \hline
 Plurality      & \cmark & \cmark & \xmark & \xmark   \\
 Anti-Plurality & \cmark & \cmark & \xmark & \xmark   \\
 Borda Count   & \cmark & \cmark & \cmark & \xmark \\
 Dowdall       & \cmark & \cmark & \cmark & \xmark   \\
 Top Two       & \cmark & \cmark & \xmark & \xmark   \\
 Hare  & \cmark & \cmark & \xmark & \xmark   \\
 Coombs Method  & \cmark &  \cmark & \cmark & \xmark   \\
 Baldwin  & \cmark &  \cmark & \cmark & \xmark  \\
 Nanson & \cmark & \cmark & \xmark & \xmark   \\
 Pairwise Majority  & \cmark & \xmark & \cmark & \cmark  \\
 Copeland  & \cmark & \cmark & \cmark & \xmark  \\
 Minimax  & \cmark & \cmark  & \xmark & \xmark  \\
 Ranked Pairs & \cmark & \cmark & \cmark & \xmark\\
\end{tabular}
\end{table}

\noindent For a formal proof of most of the above results, see \cite{robinson2010mathematics}.

\section{Simulations}
All computational models were implemented using C++ and Python. The voting systems were tested against Arrow's criterion for three to six alternatives.  All data tables
can be found in Appendix A.

Two preference distributions, Impartial Culture and Impartial Anonymous Culture \parencite{cultures} are considered. Impartial Culture (IC) assumes each voter is equally likely to select any of the possible strict preference orders. Impartial Anonymous Culture (IAC) assumes each preference schedule is equally likely to occur. All pseudo-random number generation was done using the Mersenne Twister engine \parencite{MTwist} in C++ random library.  

For each setting (fixing the criterion, number of voters, number of alternatives, and culture) 10,000 preference schedules were generated. Then the violation frequency, the percentage of times a voting system violates the criterion, was calculated for each of the 13 voting systems. The joint-violation frequencies are computed by calculating the frequency with which systems violate any one of Arrow's criteria. Due to the simulation process, the frequency of violations calculated are numerical probabilities, not theoretical. 

All computations were carried out at the Advanced Research Computing at Hopkins (ARCH) core facility  (rockfish.jhu.edu), which is supported by the National Science Foundation (NSF) grant number OAC1920103.

\section{Results and Analysis}
The following sections detail the likelihood of voting systems violating Arrow's social choice criterion. As preference schedules are drawn from IC and IAC uniform distributions, UD is satisfied by assumption. The analysis for the remaining criteria is given below. ND is excluded as all 13 voting systems satisfy ND. 

\subsection{Independence of Irrelevant Alternatives (IIA)}

The IIA algorithm in Dougherty and Heckelman (2020) works for three-alternative elections. We use a randomized IIA algorithm that enables analysis for elections with any number of alternatives. Our algorithm finds higher number of violations in most cases. The differences in the algorithms are noted in Appendix B and the code is linked in the Declarations section of the paper.  

\textbf{Key Observations}
\begin{enumerate}

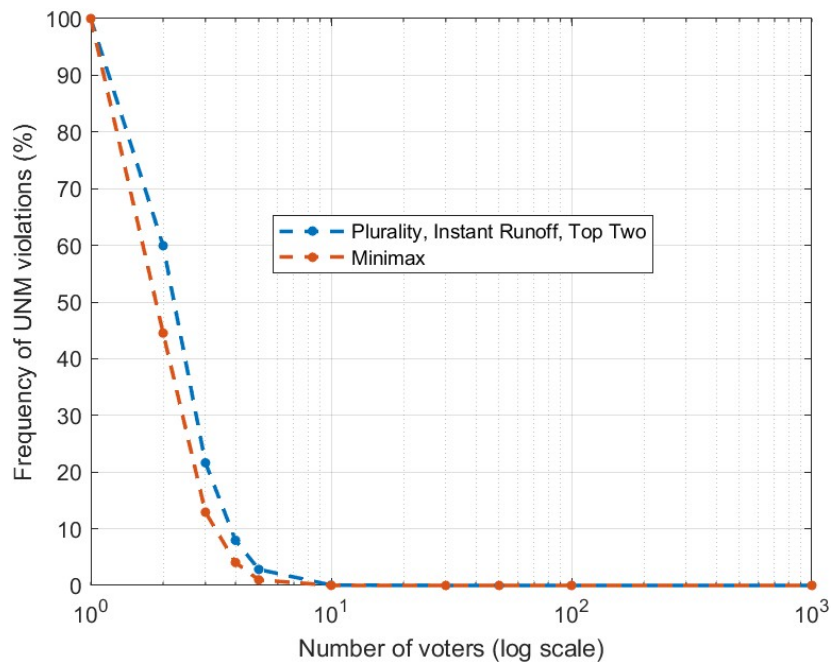
    \item For a fixed number of voters, as the number of alternatives increases, IIA violations increase. In elections with four alternatives, violation frequencies are greater than 89\% under both IC and IAC. For elections with at least five alternatives, IIA is almost always violated.

    \item For a fixed number of alternatives, as the number of voters increases, IIA violations approach 100\% under IC. 
    
    \item In alignment with Theorem 1 in \cite{Arrow}, for at least three alternatives and two voters, Group 1 systems violate IIA 100\%.  
\end{enumerate}

\noindent Tables 2 and 3 show percentage violations for three and four alternative elections under IAC. 

\begin{table}[!ht]
  \centering
  \caption{Percentage IIA Violations for 3 Alternatives (IAC Data)}
    \begin{tabular}{|l|rrrrrr|}
    \hline
    \multicolumn{1}{|c|}{\multirow{2}[4]{*}{Voting System}} & \multicolumn{6}{c|}{Number of Voters} \\
\cline{2-7}       & 3     & 4       & 10    & 30       & 100   & 10000  \\
    \hline
Group 1      & 100.00   & 100.00   & 100.00   & 100.00   & 100.00   & 100.00  \\
Copeland       & 89.08 & 56.66 & 97.75 & 99.80  & 100.00   & 99.98 \\
Dowdall        & 89.08 & 95.03  & 91.8  & 96.72 &  96.91 & 96.93 \\
Borda    & 89.08 & 56.66 & 74.29 & 84.52 & 84.14 & 84.12 \\
Coombs         & 89.08 & 56.66  & 74.29 & 87.4  &  85.7  & 85.91 \\
Baldwin        & 89.08 & 56.66  & 74.29 & 83.23 &  83.35 & 83.14
\\
Ranked Pairs & 50.32 & 56.66 & 71.64 & 75.24 & 75.97 & 75.87 \\

    \hline
    \end{tabular}%
  \label{tab:addlabel2}%
\end{table}%

\begin{table}[H]
  \centering
  \caption{Percentage IIA Violations for 4 Alternatives (IAC Data)}
    \begin{tabular}{|l|rrrrrr|}
    \hline
    \multicolumn{1}{|c|}{\multirow{2}[4]{*}{Voting System}} & \multicolumn{6}{c|}{Number of Voters} \\
\cline{2-7}                  & 3     & 4     & 10    & 30     & 100   & 10000 \\
    \hline
Group 1     & 100.00 & 100.00 & 100.00 & 100.00 & 100.00 & 100.00 \\
Copeland    & 99.08  & 99.81  & 100.00 & 100.00 & 100.00 & 100.00 \\
Dowdall     & 99.08  & 99.81  & 100.00 & 100.00 & 100.00 & 100.00 \\
Borda       & 99.08  & 99.81  & 99.90  & 99.87  & 99.93  & 99.90  \\
Coombs      & 99.08  & 99.81  & 99.91  & 100.00 & 100.00 & 99.99  \\
Baldwin     & 99.08  & 99.81  & 99.76  & 99.76  & 99.75  & 99.74  \\
Ranked Pairs & 89.86  & 89.77  & 97.23  & 98.54  & 98.51  & 98.66 \\

    \hline
    \end{tabular}%
  \label{tab:addlabel3}%
\end{table}%

Among the voting systems compared, Ranked Pairs violates IIA least frequently. For elections with 10,000 voters and 3 alternatives, Ranked Pairs violates IIA at a frequency of 75.87\% for 10000 voters. For the same election, Coombs, Borda, and Baldwin all violate IIA at frequencies between 83\% and 86\%.

Our three-alternative data substantially differs from Dougherty and Heckelman (2020). For systems not in Group 1, our algorithm reports greater percentage IIA violations. For Coombs and Copeland, the difference in the IAC data is around 20 percentage points. These differences are further explained in Appendix B. 

\subsection{Pareto Unanimity (U)}
A \textit{Pareto pair} is defined to be a pair of alternatives where all voters prefer one alternative in the pair over the other.

Of the systems considered, only Group 1 systems violate U. Even for these systems, U is nearly always satisfied. In elections with at least 10 voters, all systems have less than a 1\% probability of violating U. As the number of voters increases, the frequency of violations decreases because Pareto pairs become less likely. Table 4 shows the frequency with which systems violate U for elections with 3 voters.

\begin{table}[!ht]
    \caption{Percentage Unanimity Violations for 3 Voters (IAC)}

    \centering

\begin{tabular}{|r|rrrr|}
\hline
 \multirow{2}{5em}{Voting System} & \multicolumn{4}{c|}{Number of alternatives}  \\
\cline{2-5}
      & 3     & 4     & 5     & 6 \\
\hline
Plurality      & 10.92 & 22.35 & 40.91 & 62.69 \\
Hare & 10.92 & 22.35 & 40.91 & 62.69 \\
Top Two        & 10.92 & 22.35 & 40.91 & 62.69 \\
Anti-Plurality & 10.92 & 22.27 & 40.76 & 61.96 \\
Minimax        & 10.92 & 13.07 & 18.94 & 29.31 \\
Nanson         & 10.92 & 7.58  & 20.27 & 20.64 \\
\hline
\end{tabular}%

\end{table}

 Although percentage U violations are very low in most cases, they are common in elections with many alternatives and few voters. These trends are consistent with literature \parencite{CLC_sim, dougherty2011calculus}.  

The results for Hare, Plurality, and Top Two are identical. Data for all Group 1 systems is identical for three alternatives, but differ for four or more alternatives as noted in literature \parencite{Arrow}. For larger number of alternatives, Nanson and Minimax perform the best among the group, with less than 30\% violations with six alternatives.

\subsection{Transitivity (TR)}
Pairwise Majority is the only system studied that violates transitivity. Table 5 shows the percentage TR violations for Pairwise Majority. 

\textbf{Key Observations}
\begin{enumerate}
   \item For a fixed number of voters, as the number of alternatives increases, transitivity violations increase. 

    \item For a fixed number of alternatives, as the number of voters increase, transitivity violations decrease and appear to converge. As the number of voters tends to infinity, the probability of violating transitivity approaches a fixed value \parencite{Gehrlein1994-GEHTEL}. For three alternative elections using IAC, that value is 6.25\%, which is consistent with our findings. 
\end{enumerate}

\begin{table}[!ht]
    \caption{Percentage Transitivity Violations Pairwise Majority}
    \begin{subtable}[t]{.5\linewidth}
        \caption{IC Data}
        \centering

\begin{tabular}{|r|rrrr|}
\hline
 \multirow{2}{5em}{Number of voters} & \multicolumn{4}{c|}{Number of alternatives}  \\
\cline{2-5}
      & 3     & 4     & 5     & 6 \\
\hline
3     & 5.54  & 17.12 & 32.60 & 48.93 \\
4     & 30.69 & 63.91 & 85.84 & 95.51 \\
10    & 23.81 & 54.73 & 78.98 & 92.50 \\
30    & 17.31 & 43.78 & 68.77 & 86.03 \\
100   & 13.08 & 35.99 & 60.08 & 78.92 \\
10000 & 9.19  & 27.36 & 48.90 & 68.60 \\
\hline
\end{tabular}%
    \end{subtable}%
   \begin{subtable}[t]{.5\linewidth}
        \centering
        \caption{IAC Data}

\begin{tabular}{|r|rrrr|}
\hline
 \multirow{2}{5em}{Number of voters} & \multicolumn{4}{c|}{Number of alternatives}  \\
\cline{2-5}      & 3     & 4     & 5     & 6 \\
\hline
3     & 3.57  & 14.97 & 31.66 & 48.87 \\
4     & 23.65 & 60.59 & 85.05 & 95.49 \\
10    & 15.50 & 48.88 & 77.53 & 92.31 \\
30    & 9.58  & 35.85 & 66.47 & 85.58 \\
100   & 7.23  & 28.51 & 56.13 & 78.15 \\
10000 & 6.34  & 24.13 & 46.72 & 67.08 \\

\hline
\end{tabular}%

    \end{subtable}
\end{table}

These results match the literature \parencite{Arrow,1968paperCP, Gehrlein1994-GEHTEL}. \footnote{For elections with at least four alternatives, the limiting value for transitivity violations using IC in our study exceeds those in Niemi and Weisberg (1968) because Niemi and Weisberg examine failures to attain a Condorcet winner while we examine transitivity violations. Bottom cycles can account for the difference.}

\subsection{Joint Violation of Arrow's Criteria}
Simulations were run to find the percentage of times a voting system violates U, IIA, and TR. The joint-violation (JV) frequencies, defined as the percentage of times a system violated at least one criterion, serve as an indication of fairness of a voting system. 

\textbf{Key Observations}
\begin{enumerate}
    \item Pairwise Majority is the least likely to violate all criteria jointly. This is due to the fact that transitivity violations are rarer than IIA violations.
    \item Of the systems that satisfy transitivity, Ranked Pairs has the lowest JV frequencies followed by Baldwin, Borda Count, and Coombs. 
    \item  For a fixed number of voters, as the number of alternatives increases, the JV frequencies increase for every system. For elections with 5 or more alternatives, JV frequencies are almost always 100\%. 
\end{enumerate}

\begin{table}[!ht]
    \caption{Joint-Violation Frequencies for 100 voters (IAC)}

    \centering

\begin{tabular}{|r|rrrr|}
\hline
 \multirow{2}{5em}{Voting System} & \multicolumn{4}{c|}{Number of alternatives}  \\
\cline{2-5}
      & 3     & 4     & 5     & 6 \\
\hline
Group 1           & 100.00 & 100.00 & 100.00 & 100.00 \\
Copeland          & 85.50  & 100.00 & 100.00 & 100.00 \\
Dowdall           & 96.84  & 100.00 & 100.00 & 100.00 \\
Borda             & 83.85  & 99.94  & 100.00 & 100.00 \\
Coombs            & 85.50  & 100.00 & 100.00 & 100.00 \\
Baldwin           & 82.86  & 99.84  & 100.00 & 100.00 \\
Ranked Pairs      & 75.66  & 98.67  & 100.00 & 100.00 \\
Pairwise Majority & 7.23   & 28.51  & 56.13  & 78.15  \\
\hline
\end{tabular}%

\end{table}

The conclusion for three-alternative elections is consistent with literature. Dougherty and Heckelman (2020) agrees that Pairwise Majority has the lowest JV frequencies. However, because differences in the IIA data, which is discussed in Appendix B, results for Copeland, Coombs, Dowdall, Borda, Ranked Pairs, and Baldwin do not match exactly. These systems have higher JV frequencies than in the 2020 study.

\section{Conclusion}
This study contributes to the enduring challenge in social choice theory of identifying voting systems that perform best with respect to Arrow’s axiomatic criteria. While Arrow’s impossibility theorem establishes that no voting system can simultaneously satisfy all fairness conditions under general preference domains, our results provide an empirical assessment of how often different systems fail in practice. Extending prior work by Dougherty and Heckelman (2020), our simulations consider larger electorates and a wider range of alternatives, offering a more comprehensive picture of violation frequencies across commonly used preference aggregation rules.

Our central finding - that Pairwise Majority exhibits the lowest joint-violation frequency among all tested systems - reinforces prior theoretical observations (Gehrlein, 1994; Niemi \& Weisberg, 1968) and suggests that even within the constraints of Arrow’s theorem, some systems may better approximate normative ideals in practical settings. Importantly, we demonstrate that the likelihood of IIA violations rises sharply with the number of alternatives, a trend consistent with both theoretical expectations and empirical findings (Nurmi, 1983; Börgers, 2010). Since almost every voting rule is susceptible to IIA violations, this result highlights their ubiquity and underscores that systems must often be judged on other normative grounds unless pairwise majority rule is employed.

These insights are especially relevant in contemporary democratic societies where elections increasingly involve crowded fields and diverse voter preferences. At the same time, we caution that our conclusions are based on simulations under the uniform distribution of preferences; different distributions, particularly skewed ones, may yield different violation probabilities.

The implications of these results extend beyond theoretical interest. As public debate over electoral reform intensifies, a clearer understanding of the probabilistic fairness of voting systems is essential. While perfect fairness is unattainable, the empirical identification of systems with relatively lower joint violation frequencies, such as Pairwise Majority, can inform institutional design. Moreover, our enhanced methodology for detecting IIA violations offers more precise estimates, thereby addressing discrepancies in previous computational studies.

Our findings help show that while Arrow's theorem remains an immutable constraint, its practical implications are context-dependent and, to some extent, quantifiable. Future work might extend this analysis to include strategic behavior, correlated preferences, or alternative fairness frameworks (e.g., monotonicity, Condorcet consistency), continuing the search for voting systems that, while not flawless, are less prone to failure in the eyes of theory and society alike.

\section{Declarations}
\textbf{Competing Interests}
The authors declare that they have no known competing financial interests or personal relationships that could have appeared to influence the work reported in this paper.

\noindent
\textbf{Data Availability Statement}
The raw data generated during this study is available upon request. All relevant processed data is included within this article and the appendix. The Python code is avaliable to run here: https://github.com/review123-coder/VotingSystemAnonymous. 

\newpage
\printbibliography

\newpage
\appendix
\section{Data Tables}
\subsection{IIA}

\begin{table}[H]
    \caption{Percentage IIA Violations Group 1}
    \begin{subtable}[t]{.5\linewidth}
        \caption{IC Data}
        \centering
\begin{tabular}{|r|rrrr|}
\hline
 \multirow{2}{5em}{Number of voters} & \multicolumn{4}{c|}{Number of alternatives}  \\
\cline{2-5}      & 3     & 4     & 5     & 6 \\
\hline
3     & 100.00 & 100.00 & 100.00 & 100.00 \\
4     & 100.00 & 100.00 & 100.00 & 100.00 \\
10    & 100.00 & 100.00 & 100.00 & 100.00 \\
30    & 100.00 & 100.00 & 100.00 & 100.00 \\
100   & 100.00 & 100.00 & 100.00 & 100.00 \\
10000 & 100.00 & 100.00 & 100.00 & 100.00 \\
\hline
\end{tabular}%
        
        \end{subtable}%
   \begin{subtable}[t]{.5\linewidth}
        \centering
        \caption{IAC Data}

\begin{tabular}{|r|rrrr|}
\hline
 \multirow{2}{5em}{Number of voters} & \multicolumn{4}{c|}{Number of alternatives}  \\
\cline{2-5}      & 3     & 4     & 5     & 6 \\
\hline
3     & 100.00 & 100.00 & 100.00 & 100.00 \\
4     & 100.00 & 100.00 & 100.00 & 100.00 \\
10    & 100.00 & 100.00 & 100.00 & 100.00 \\
30    & 100.00 & 100.00 & 100.00 & 100.00 \\
100   & 100.00 & 100.00 & 100.00 & 100.00 \\
10000 & 100.00 & 100.00 & 100.00 & 100.00 \\
\hline
\end{tabular}%

         \end{subtable}
\end{table}


\begin{table}[H]
    \caption{Percentage IIA Violations Borda}
    \begin{subtable}[t]{.5\linewidth}
        \caption{IC Data}
        \centering

\begin{tabular}{|r|rrrr|}
\hline
 \multirow{2}{5em}{Number of voters} & \multicolumn{4}{c|}{Number of alternatives}  \\
\cline{2-5}      & 3     & 4     & 5     & 6 \\
\hline
3     & 97.35  & 99.78  & 99.98  & 100.00 \\
4     & 73.63  & 99.96  & 100.00 & 100.00 \\
10    & 94.79  & 100.00 & 100.00 & 100.00 \\
30    & 99.97  & 100.00 & 100.00 & 100.00 \\
100   & 100.00 & 100.00 & 100.00 & 100.00 \\
10000 & 100.00 & 100.00 & 100.00 & 100.00 \\
\hline
\end{tabular}%

        \end{subtable}%
   \begin{subtable}[t]{.5\linewidth}
        \centering
        \caption{IAC Data}

\begin{tabular}{|r|rrrr|}
\hline
 \multirow{2}{5em}{Number of voters} & \multicolumn{4}{c|}{Number of alternatives}  \\
\cline{2-5}      & 3     & 4     & 5     & 6 \\
\hline
3     & 89.08  & 99.08  & 99.96  & 100.00 \\
4     & 56.66  & 99.81  & 100.00 & 100.00 \\
10    & 74.29  & 99.90  & 100.00 & 100.00 \\
30    & 84.52  & 99.87  & 100.00 & 100.00 \\
100   & 84.14  & 99.93  & 100.00 & 100.00 \\
10000 & 84.12  & 99.90  & 100.00 & 100.00 \\
\hline
\end{tabular}%

         \end{subtable}
\end{table}


\begin{table}[H]
    \caption{Percentage IIA Violations Coombs}
    \begin{subtable}[t]{.5\linewidth}
        \caption{IC Data}
        \centering

\begin{tabular}{|r|rrrr|}
\hline
 \multirow{2}{5em}{Number of voters} & \multicolumn{4}{c|}{Number of alternatives}  \\
\cline{2-5}      & 3     & 4     & 5     & 6 \\
\hline
3     & 97.35  & 99.78  & 99.98  & 100.00 \\
4     & 73.63  & 99.96  & 100.00 & 100.00 \\
10    & 94.79  & 100.00 & 100.00 & 100.00 \\
30    & 99.99  & 100.00 & 100.00 & 100.00 \\
100   & 100.00 & 100.00 & 100.00 & 100.00 \\
10000 & 100.00 & 100.00 & 100.00 & 100.00 \\
\hline
\end{tabular}%

        \end{subtable}%
   \begin{subtable}[t]{.5\linewidth}
        \centering
        \caption{IAC Data}

\begin{tabular}{|r|rrrr|}
\hline
 \multirow{2}{5em}{Number of voters} & \multicolumn{4}{c|}{Number of alternatives}  \\
\cline{2-5}      & 3     & 4     & 5     & 6 \\
\hline
3     & 89.08 & 99.08  & 99.96  & 100.00 \\
4     & 56.66 & 99.81  & 100.00 & 100.00 \\
10    & 74.29 & 99.91  & 100.00 & 100.00 \\
30    & 87.40 & 100.00 & 100.00 & 100.00 \\
100   & 85.70 & 100.00 & 100.00 & 100.00 \\
10000 & 85.91 & 99.99  & 100.00 & 100.00 \\
\hline
\end{tabular}%

         \end{subtable}
\end{table}

\begin{table}[H]
    \caption{Percentage IIA Violations Dowdall}
    \begin{subtable}[t]{.5\linewidth}
        \caption{IC Data}
        \centering
\begin{tabular}{|r|rrrr|}
\hline
 \multirow{2}{5em}{Number of voters} & \multicolumn{4}{c|}{Number of alternatives}  \\
\cline{2-5}      & 3     & 4     & 5     & 6 \\
\hline
3     & 97.35  & 99.78  & 99.98  & 100.00 \\
4     & 99.52  & 99.96  & 100.00 & 100.00 \\
10    & 99.79  & 100.00 & 100.00 & 100.00 \\
30    & 100.00 & 100.00 & 100.00 & 100.00 \\
100   & 100.00 & 100.00 & 100.00 & 100.00 \\
10000 & 100.00 & 100.00 & 100.00 & 100.00 \\
\hline
\end{tabular}%

        \end{subtable}%
   \begin{subtable}[t]{.5\linewidth}
        \centering
        \caption{IAC Data}

\begin{tabular}{|r|rrrr|}
\hline
 \multirow{2}{5em}{Number of voters} & \multicolumn{4}{c|}{Number of alternatives}  \\
\cline{2-5}      & 3     & 4     & 5     & 6 \\
\hline
3     & 89.08 & 99.08  & 99.96  & 100.00 \\
4     & 95.03 & 99.81  & 100.00 & 100.00 \\
10    & 91.80 & 100.00 & 100.00 & 100.00 \\
30    & 96.72 & 100.00 & 100.00 & 100.00 \\
100   & 96.91 & 100.00 & 100.00 & 100.00 \\
10000 & 96.93 & 100.00 & 100.00 & 100.00 \\
\hline
\end{tabular}%
         \end{subtable}
\end{table}


\begin{table}[H]
    \caption{Percentage IIA Violations Copeland}
    \begin{subtable}[t]{.5\linewidth}
        \caption{IC Data}
        \centering

\begin{tabular}{|r|rrrr|}
\hline
 \multirow{2}{5em}{Number of voters} & \multicolumn{4}{c|}{Number of alternatives}  \\
\cline{2-5}      & \multicolumn{1}{r|}{3} & 4     & 5     & 6 \\
\hline
3     & 97.35  & 99.78  & 99.98  & 100.00 \\
4     & 73.63  & 99.96  & 100.00 & 100.00 \\
10    & 99.99  & 100.00 & 100.00 & 100.00 \\
30    & 100.00 & 100.00 & 100.00 & 100.00 \\
100   & 100.00 & 100.00 & 100.00 & 100.00 \\
10000 & 100.00 & 100.00 & 100.00 & 100.00 \\

\hline
\end{tabular}%
        \end{subtable}%
   \begin{subtable}[t]{.5\linewidth}
        \centering
        \caption{IAC Data}

\begin{tabular}{|r|rrrr|}
\hline
 \multirow{2}{5em}{Number of voters} & \multicolumn{4}{c|}{Number of alternatives}  \\
\cline{2-5}      & \multicolumn{1}{r|}{3} & 4     & 5     & 6 \\
\hline
3     & 89.08  & 99.08  & 99.96  & 100.00 \\
4     & 56.66  & 99.81  & 100.00 & 100.00 \\
10    & 97.75  & 100.00 & 100.00 & 100.00 \\
30    & 99.80  & 100.00 & 100.00 & 100.00 \\
100   & 100.00 & 100.00 & 100.00 & 100.00 \\
10000 & 99.98  & 100.00 & 100.00 & 100.00 \\
\hline
\end{tabular}%

         \end{subtable}
\end{table}

\begin{table}[H]
    \caption{Percentage IIA Violations Baldwin}
    \begin{subtable}[t]{.5\linewidth}
        \caption{IC Data}
        \centering

\begin{tabular}{|r|rrrr|}
\hline
 \multirow{2}{5em}{Number of voters} & \multicolumn{4}{c|}{Number of alternatives}  \\
\cline{2-5}      & 3     & 4     & 5     & 6 \\
\hline
3     & 97.35  & 99.78  & 99.98  & 100.00 \\
4     & 73.63  & 99.96  & 100.00 & 100.00 \\
10    & 94.79  & 99.99  & 100.00 & 100.00 \\
30    & 99.96  & 100.00 & 100.00 & 100.00 \\
100   & 100.00 & 100.00 & 100.00 & 100.00 \\
10000 & 100.00 & 100.00 & 100.00 & 100.00 \\
\hline
\end{tabular}%

        \end{subtable}%
   \begin{subtable}[t]{.5\linewidth}
        \centering
        \caption{IAC Data}

\begin{tabular}{|r|rrrr|}
\hline
 \multirow{2}{5em}{Number of voters} & \multicolumn{4}{c|}{Number of alternatives}  \\
\cline{2-5}      & 3     & 4     & 5     & 6 \\
\hline
3     & 89.08 & 99.08 & 99.96  & 100.00 \\
4     & 56.66 & 99.81 & 100.00 & 100.00 \\
10    & 74.29 & 99.76 & 100.00 & 100.00 \\
30    & 83.23 & 99.76 & 100.00 & 100.00 \\
100   & 83.35 & 99.75 & 100.00 & 100.00 \\
10000 & 83.14 & 99.74 & 100.00 & 100.00 \\
\hline
\end{tabular}%
         \end{subtable}
\end{table}


\begin{table}[H]
    \caption{Percentage IIA Violations Ranked Pairs}
    \begin{subtable}[t]{.5\linewidth}
        \caption{IC Data}
        \centering

\begin{tabular}{|r|rrrr|}
\hline
 \multirow{2}{5em}{Number of voters} & \multicolumn{4}{c|}{Number of alternatives}  \\
\cline{2-5}      & 3     & 4     & 5     & 6 \\
\hline
3     & 58.43  & 92.39  & 99.47  & 99.93  \\
4     & 73.63  & 93.71  & 99.43  & 99.98  \\
10    & 92.88  & 99.48  & 100.00 & 100.00 \\
30    & 99.32  & 99.94  & 100.00 & 100.00 \\
100   & 100.00 & 100.00 & 100.00 & 100.00 \\
10000 & 100.00 & 100.00 & 100.00 & 100.00 \\
\hline
\end{tabular}%

        \end{subtable}%
   \begin{subtable}[t]{.5\linewidth}
        \centering
        \caption{IAC Data}

\begin{tabular}{|r|rrrr|}
\hline
 \multirow{2}{5em}{Number of voters} & \multicolumn{4}{c|}{Number of alternatives}  \\
\cline{2-5}      & 3     & 4     & 5     & 6 \\
\hline
3     & 50.32 & 89.86 & 99.26  & 99.99  \\
4     & 56.66 & 89.77 & 99.26  & 99.95  \\
10    & 71.64 & 97.23 & 99.92  & 100.00 \\
30    & 75.24 & 98.54 & 99.99  & 100.00 \\
100   & 75.97 & 98.51 & 100.00 & 100.00 \\
10000 & 75.87 & 98.66 & 100.00 & 100.00 \\

\hline
\end{tabular}%
         \end{subtable}
\end{table}


\subsection{Unanimity}

\begin{table}[H]
    \caption{Percentage U Violations Plurality, Top Two, Hare}
    \begin{subtable}[t]{.5\linewidth}
        \caption{IC Data}
        \centering

\begin{tabular}{|r|rrrr|}
\hline
 \multirow{2}{5em}{Number of voters} & \multicolumn{4}{c|}{Number of alternatives}  \\
\cline{2-5}      & 3     & 4     & 5     & 6 \\
\hline
3     & 2.65 & 17.88 & 39.84 & 63.07 \\
4     & 0.48 & 4.73  & 14.22 & 27.71 \\
10    & 0.00 & 0.00  & 0.01  & 0.04  \\
30    & 0.00 & 0.00  & 0.00  & 0.00  \\
100   & 0.00 & 0.00  & 0.00  & 0.00  \\
10000 & 0.00 & 0.00  & 0.00  & 0.00 \\
\hline
\end{tabular}%

        \end{subtable}%
   \begin{subtable}[t]{.5\linewidth}
        \centering
        \caption{IAC Data}

\begin{tabular}{|r|rrrr|}
\hline
 \multirow{2}{5em}{Number of voters} & \multicolumn{4}{c|}{Number of alternatives}  \\
\cline{2-5}      & 3     & 4     & 5     & 6 \\
\hline
3    & 10.92 & 22.35 & 40.91 & 62.69 \\
4    & 4.97  & 7.87  & 15.37 & 28.00 \\
10   & 0.22  & 0.06  & 0.01  & 0.08  \\
30   & 0.00  & 0.00  & 0.00  & 0.00  \\
100  & 0.00  & 0.00  & 0.00  & 0.00  \\
1000 & 0.00  & 0.00  & 0.00  & 0.00 \\

\hline
\end{tabular}%
         \end{subtable}
\end{table}

\begin{table}[H]
    \caption{Percentage U Violations Minimax}
    \begin{subtable}[t]{.5\linewidth}
        \caption{IC Data}
        \centering

\begin{tabular}{|r|rrrr|}
\hline
 \multirow{2}{5em}{Number of voters} & \multicolumn{4}{c|}{Number of alternatives}  \\
\cline{2-5}      & 3     & 4     & 5     & 6 \\
\hline
3     & 2.65 & 9.25 & 18.01 & 29.83 \\
4     & 0.48 & 1.71 & 3.73  & 6.68  \\
10    & 0.00 & 0.00 & 0.00  & 0.00  \\
30    & 0.00 & 0.00 & 0.00  & 0.00  \\
100   & 0.00 & 0.00 & 0.00  & 0.00  \\
10000 & 0.00 & 0.00 & 0.00  & 0.00 \\
\hline
\end{tabular}%

        \end{subtable}%
   \begin{subtable}[t]{.5\linewidth}
        \centering
        \caption{IAC Data}

\begin{tabular}{|r|rrrr|}
\hline
 \multirow{2}{5em}{Number of voters} & \multicolumn{4}{c|}{Number of alternatives}  \\
\cline{2-5}      & 3     & 4     & 5     & 6 \\
\hline
3    & 10.92 & 13.07 & 18.94 & 29.31 \\
4    & 4.97  & 3.81  & 4.16  & 7.18  \\
10   & 0.22  & 0.02  & 0.00  & 0.00  \\
30   & 0.00  & 0.00  & 0.00  & 0.00  \\
100  & 0.00  & 0.00  & 0.00  & 0.00  \\
1000 & 0.00  & 0.00  & 0.00  & 0.00 \\

\hline
\end{tabular}%
         \end{subtable}
\end{table}

\begin{table}[H]
    \caption{Percentage U Violations Anti-Plurality}
    \begin{subtable}[t]{.5\linewidth}
        \caption{IC Data}
        \centering

\begin{tabular}{|r|rrrr|}
\hline
 \multirow{2}{5em}{Number of voters} & \multicolumn{4}{c|}{Number of alternatives}  \\
\cline{2-5}      & 3     & 4     & 5     & 6 \\
\hline
3     & 2.65 & 17.20 & 40.29 & 62.79 \\
4     & 0.48 & 4.82  & 14.19 & 28.51 \\
10    & 0.00 & 0.00  & 0.03  & 0.04  \\
30    & 0.00 & 0.00  & 0.00  & 0.00  \\
100   & 0.00 & 0.00  & 0.00  & 0.00  \\
10000 & 0.00 & 0.00  & 0.00  & 0.00 \\
\hline
\end{tabular}%

        \end{subtable}%
   \begin{subtable}[t]{.5\linewidth}
        \centering
        \caption{IAC Data}

\begin{tabular}{|r|rrrr|}
\hline
 \multirow{2}{5em}{Number of voters} & \multicolumn{4}{c|}{Number of alternatives}  \\
\cline{2-5}      & 3     & 4     & 5     & 6 \\
\hline
3    & 10.92 & 22.27 & 40.76 & 61.96 \\
4    & 4.97  & 7.98  & 15.12 & 28.12 \\
10   & 0.22  & 0.03  & 0.02  & 0.04  \\
30   & 0.00  & 0.00  & 0.00  & 0.00  \\
100  & 0.00  & 0.00  & 0.00  & 0.00  \\
1000 & 0.00  & 0.00  & 0.00  & 0.00 \\

\hline
\end{tabular}%
         \end{subtable}
\end{table}

\begin{table}[H]
    \caption{Percentage U Violations Nanson}
    \begin{subtable}[t]{.5\linewidth}
        \caption{IC Data}
        \centering

\begin{tabular}{|r|rrrr|}
\hline
 \multirow{2}{5em}{Number of voters} & \multicolumn{4}{c|}{Number of alternatives}  \\
\cline{2-5}      & 3     & 4     & 5     & 6 \\
\hline
3     & 2.65 & 5.09 & 19.27 & 21.60 \\
4     & 0.48 & 2.70 & 5.63  & 9.50  \\
10    & 0.00 & 0.00 & 0.00  & 0.01  \\
30    & 0.00 & 0.00 & 0.00  & 0.00  \\
100   & 0.00 & 0.00 & 0.00  & 0.00  \\
10000 & 0.00 & 0.00 & 0.00  & 0.00 \\
\hline
\end{tabular}%

        \end{subtable}%
   \begin{subtable}[t]{.5\linewidth}
        \centering
        \caption{IAC Data}

\begin{tabular}{|r|rrrr|}
\hline
 \multirow{2}{5em}{Number of voters} & \multicolumn{4}{c|}{Number of alternatives}  \\
\cline{2-5}      & 3     & 4     & 5     & 6 \\
\hline
3    & 10.92 & 7.58 & 20.27 & 20.64 \\
4    & 4.97  & 5.20 & 6.25  & 9.43  \\
10   & 0.22  & 0.05 & 0.02  & 0.02  \\
30   & 0.00  & 0.00 & 0.00  & 0.00  \\
100  & 0.00  & 0.00 & 0.00  & 0.00  \\
1000 & 0.00  & 0.00 & 0.00  & 0.00 \\

\hline
\end{tabular}%
         \end{subtable}
\end{table}

\subsection{Joint Violation Frequencies}

\begin{table}[H]
    \caption{JV Frequencies Group 1}
    \begin{subtable}[t]{.5\linewidth}
        \caption{IC Data}
        \centering

\begin{tabular}{|r|rrrr|}
\hline
 \multirow{2}{5em}{Number of voters} & \multicolumn{4}{c|}{Number of alternatives}  \\
\cline{2-5}      & 3     & 4     & 5     & 6 \\
\hline
3     & 100.00 & 100.00 & 100.00 & 100.00 \\
4     & 100.00 & 100.00 & 100.00 & 100.00 \\
10    & 100.00 & 100.00 & 100.00 & 100.00 \\
30    & 100.00 & 100.00 & 100.00 & 100.00 \\
100   & 100.00 & 100.00 & 100.00 & 100.00 \\
10000 & 100.00 & 100.00 & 100.00 & 100.00 \\
\hline
\end{tabular}%

        \end{subtable}%
   \begin{subtable}[t]{.5\linewidth}
        \centering
        \caption{IAC Data}

\begin{tabular}{|r|rrrr|}
\hline
 \multirow{2}{5em}{Number of voters} & \multicolumn{4}{c|}{Number of alternatives}  \\
\cline{2-5}      & 3     & 4     & 5     & 6 \\
\hline
3     & 100.00 & 100.00 & 100.00 & 100.00 \\
4     & 100.00 & 100.00 & 100.00 & 100.00 \\
10    & 100.00 & 100.00 & 100.00 & 100.00 \\
30    & 100.00 & 100.00 & 100.00 & 100.00 \\
100   & 100.00 & 100.00 & 100.00 & 100.00 \\
10000 & 100.00 & 100.00 & 100.00 & 100.00 \\

\hline
\end{tabular}%
         \end{subtable}
\end{table}

\begin{table}[H]
    \caption{JV Frequencies Borda Count}
    \begin{subtable}[t]{.5\linewidth}
        \caption{IC Data}
        \centering

\begin{tabular}{|r|rrrr|}
\hline
 \multirow{2}{5em}{Number of voters} & \multicolumn{4}{c|}{Number of alternatives}  \\
\cline{2-5}      & 3     & 4     & 5     & 6 \\
\hline
3     & 97.30  & 99.87  & 100.00 & 100.00 \\
4     & 73.42  & 100.00 & 100.00 & 100.00 \\
10    & 94.84  & 99.99  & 100.00 & 100.00 \\
30    & 99.89  & 100.00 & 100.00 & 100.00 \\
100   & 100.00 & 100.00 & 100.00 & 100.00 \\
10000 & 100.00 & 100.00 & 100.00 & 100.00 \\
\hline
\end{tabular}%

        \end{subtable}%
   \begin{subtable}[t]{.5\linewidth}
        \centering
        \caption{IAC Data}

\begin{tabular}{|r|rrrr|}
\hline
 \multirow{2}{5em}{Number of voters} & \multicolumn{4}{c|}{Number of alternatives}  \\
\cline{2-5}      & 3     & 4     & 5     & 6 \\
\hline
3     & 89.38 & 99.17 & 99.92  & 100.00 \\
4     & 55.94 & 99.81 & 99.99  & 100.00 \\
10    & 74.84 & 99.88 & 100.00 & 100.00 \\
30    & 84.52 & 99.93 & 100.00 & 100.00 \\
100   & 83.85 & 99.94 & 100.00 & 100.00 \\
10000 & 84.63 & 99.85 & 100.00 & 100.00 \\

\hline
\end{tabular}%
         \end{subtable}
\end{table}

\begin{table}[H]
    \caption{JV Frequencies Coombs}
    \begin{subtable}[t]{.5\linewidth}
        \caption{IC Data}
        \centering

\begin{tabular}{|r|rrrr|}
\hline
 \multirow{2}{5em}{Number of voters} & \multicolumn{4}{c|}{Number of alternatives}  \\
\cline{2-5}      & 3     & 4     & 5     & 6 \\
\hline
3     & 97.30  & 99.87  & 100.00 & 100.00 \\
4     & 73.42  & 100.00 & 100.00 & 100.00 \\
10    & 94.84  & 99.99  & 100.00 & 100.00 \\
30    & 99.98  & 100.00 & 100.00 & 100.00 \\
100   & 100.00 & 100.00 & 100.00 & 100.00 \\
10000 & 100.00 & 100.00 & 100.00 & 100.00 \\
\hline
\end{tabular}%

        \end{subtable}%
   \begin{subtable}[t]{.5\linewidth}
        \centering
        \caption{IAC Data}

\begin{tabular}{|r|rrrr|}
\hline
 \multirow{2}{5em}{Number of voters} & \multicolumn{4}{c|}{Number of alternatives}  \\
\cline{2-5}      & 3     & 4     & 5     & 6 \\
\hline
3     & 89.38 & 99.17  & 99.92  & 100.00 \\
4     & 55.94 & 99.81  & 99.99  & 100.00 \\
10    & 74.84 & 99.90  & 100.00 & 100.00 \\
30    & 87.19 & 99.99  & 100.00 & 100.00 \\
100   & 85.50 & 100.00 & 100.00 & 100.00 \\
10000 & 86.60 & 99.99  & 100.00 & 100.00 \\

\hline
\end{tabular}%
         \end{subtable}
\end{table}

\begin{table}[H]
    \caption{JV Frequencies Dowdall}
    \begin{subtable}[t]{.5\linewidth}
        \caption{IC Data}
        \centering

\begin{tabular}{|r|rrrr|}
\hline
 \multirow{2}{5em}{Number of voters} & \multicolumn{4}{c|}{Number of alternatives}  \\
\cline{2-5}      & 3     & 4     & 5     & 6 \\
\hline
3     & 97.30  & 99.87  & 100.00 & 100.00 \\
4     & 99.41  & 100.00 & 100.00 & 100.00 \\
10    & 99.78  & 100.00 & 100.00 & 100.00 \\
30    & 100.00 & 100.00 & 100.00 & 100.00 \\
100   & 100.00 & 100.00 & 100.00 & 100.00 \\
10000 & 100.00 & 100.00 & 100.00 & 100.00 \\
\hline
\end{tabular}%

        \end{subtable}%
   \begin{subtable}[t]{.5\linewidth}
        \centering
        \caption{IAC Data}

\begin{tabular}{|r|rrrr|}
\hline
 \multirow{2}{5em}{Number of voters} & \multicolumn{4}{c|}{Number of alternatives}  \\
\cline{2-5}      & 3     & 4     & 5     & 6 \\
\hline
3     & 89.38 & 99.17  & 99.92  & 100.00 \\
4     & 94.94 & 99.81  & 99.99  & 100.00 \\
10    & 92.59 & 99.98  & 100.00 & 100.00 \\
30    & 96.70 & 100.00 & 100.00 & 100.00 \\
100   & 96.84 & 100.00 & 100.00 & 100.00 \\
10000 & 97.03 & 100.00 & 100.00 & 100.00 \\

\hline
\end{tabular}%
         \end{subtable}
\end{table}

\begin{table}[H]
    \caption{JV Frequencies Copeland}
    \begin{subtable}[t]{.5\linewidth}
        \caption{IC Data}
        \centering

\begin{tabular}{|r|rrrr|}
\hline
 \multirow{2}{5em}{Number of voters} & \multicolumn{4}{c|}{Number of alternatives}  \\
\cline{2-5}      & 3     & 4     & 5     & 6 \\
\hline
3     & 97.30  & 99.87  & 100.00 & 100.00 \\
4     & 73.42  & 100.00 & 100.00 & 100.00 \\
10    & 100.00 & 100.00 & 100.00 & 100.00 \\
30    & 100.00 & 100.00 & 100.00 & 100.00 \\
100   & 100.00 & 100.00 & 100.00 & 100.00 \\
10000 & 100.00 & 100.00 & 100.00 & 100.00 \\
\hline
\end{tabular}%

        \end{subtable}%
   \begin{subtable}[t]{.5\linewidth}
        \centering
        \caption{IAC Data}

\begin{tabular}{|r|rrrr|}
\hline
 \multirow{2}{5em}{Number of voters} & \multicolumn{4}{c|}{Number of alternatives}  \\
\cline{2-5}      & 3     & 4     & 5     & 6 \\
\hline
3     & 89.38 & 99.17  & 99.92  & 100.00 \\
4     & 55.94 & 99.81  & 99.99  & 100.00 \\
10    & 98.10 & 100.00 & 99.99  & 100.00 \\
30    & 99.85 & 100.00 & 100.00 & 100.00 \\
100   & 99.98 & 100.00 & 100.00 & 100.00 \\
10000 & 99.97 & 100.00 & 100.00 & 100.00 \\

\hline
\end{tabular}%
         \end{subtable}
\end{table}

\begin{table}[H]
    \caption{JV Frequencies Baldwin}
    \begin{subtable}[t]{.5\linewidth}
        \caption{IC Data}
        \centering

\begin{tabular}{|r|rrrr|}
\hline
 \multirow{2}{5em}{Number of voters} & \multicolumn{4}{c|}{Number of alternatives}  \\
\cline{2-5}      & 3     & 4     & 5     & 6 \\
\hline
3     & 97.30  & 99.87  & 100.00 & 100.00 \\
4     & 73.42  & 100.00 & 100.00 & 100.00 \\
10    & 94.84  & 99.99  & 100.00 & 100.00 \\
30    & 99.99  & 100.00 & 100.00 & 100.00 \\
100   & 100.00 & 100.00 & 100.00 & 100.00 \\
10000 & 100.00 & 100.00 & 100.00 & 100.00 \\
\hline
\end{tabular}%

        \end{subtable}%
   \begin{subtable}[t]{.5\linewidth}
        \centering
        \caption{IAC Data}

\begin{tabular}{|r|rrrr|}
\hline
 \multirow{2}{5em}{Number of voters} & \multicolumn{4}{c|}{Number of alternatives}  \\
\cline{2-5}      & 3     & 4     & 5     & 6 \\
\hline
3     & 89.38 & 99.17 & 99.14  & 100.00 \\
4     & 55.94 & 99.81 & 99.10  & 100.00 \\
10    & 74.84 & 99.75 & 99.93  & 100.00 \\
30    & 83.13 & 99.73 & 100.00 & 100.00 \\
100   & 82.86 & 99.84 & 100.00 & 100.00 \\
10000 & 83.32 & 99.71 & 100.00 & 100.00 \\

\hline
\end{tabular}%
         \end{subtable}
\end{table}

\begin{table}[H]
    \caption{JV Frequencies Ranked Pairs}
    \begin{subtable}[t]{.5\linewidth}
        \caption{IC Data}
        \centering

\begin{tabular}{|r|rrrr|}
\hline
 \multirow{2}{5em}{Number of voters} & \multicolumn{4}{c|}{Number of alternatives}  \\
\cline{2-5}      & 3     & 4     & 5     & 6 \\
\hline
3     & 58.19  & 92.03  & 99.33  & 99.96  \\
4     & 73.42  & 94.06  & 99.32  & 99.98  \\
10    & 92.98  & 99.45  & 99.98  & 99.99  \\
30    & 99.34  & 99.95  & 100.00 & 100.00 \\
100   & 100.00 & 100.00 & 100.00 & 100.00 \\
10000 & 100.00 & 100.00 & 100.00 & 100.00 \\
\hline
\end{tabular}%

        \end{subtable}%
   \begin{subtable}[t]{.5\linewidth}
        \centering
        \caption{IAC Data}

\begin{tabular}{|r|rrrr|}
\hline
 \multirow{2}{5em}{Number of voters} & \multicolumn{4}{c|}{Number of alternatives}  \\
\cline{2-5}      & 3     & 4     & 5     & 6 \\
\hline
3     & 49.20 & 89.57 & 99.14  & 99.97  \\
4     & 55.94 & 89.50 & 99.10  & 99.97  \\
10    & 72.34 & 97.72 & 99.93  & 100.00 \\
30    & 74.57 & 98.29 & 100.00 & 100.00 \\
100   & 75.66 & 98.67 & 100.00 & 100.00 \\
10000 & 76.31 & 98.55 & 100.00 & 100.00 \\

\hline
\end{tabular}%
         \end{subtable}
\end{table}

\begin{table}[H]
    \caption{JV Frequencies Pairwise Majority}
    \begin{subtable}[t]{.5\linewidth}
        \caption{IC Data}
        \centering

    \begin{tabular}{|r|rrrr|}
    \hline
 \multirow{2}{5em}{Number of voters} & \multicolumn{4}{c|}{Number of alternatives}  \\
\cline{2-5}          & 3     & 4     & 5     & 6 \\
    \hline
3     & 5.54  & 17.12 & 32.60 & 48.93 \\
4     & 30.69 & 63.91 & 85.84 & 95.51 \\
10    & 23.81 & 54.73 & 78.98 & 92.50 \\
30    & 17.31 & 43.78 & 68.77 & 86.03 \\
100   & 13.08 & 35.99 & 60.08 & 78.92 \\
10000 & 9.19  & 27.36 & 48.90 & 68.60 \\
    \hline
    \end{tabular}%

        \end{subtable}%
   \begin{subtable}[t]{.5\linewidth}
        \centering
        \caption{IAC Data}

    \begin{tabular}{|r|rrrr|}
    \hline
 \multirow{2}{5em}{Number of voters} & \multicolumn{4}{c|}{Number of alternatives}  \\
\cline{2-5}          & 3     & 4     & 5     & 6 \\
    \hline
3     & 3.57  & 14.97 & 31.66 & 48.87 \\
4     & 23.65 & 60.59 & 85.05 & 95.49 \\
10    & 15.50 & 48.88 & 77.53 & 92.31 \\
30    & 9.58  & 35.85 & 66.47 & 85.58 \\
100   & 7.23  & 28.51 & 56.13 & 78.15 \\
10000 & 6.34  & 24.13 & 46.72 & 67.08 \\
    \hline
    \end{tabular}%

         \end{subtable}
\end{table}

\newpage

\section{IIA Algorithm}

The IIA algorithm used in this paper is different than that of Dougherty and Heckelman (2020) and we describe those differences here. 

As IIA is an inter-profile condition, to check if a voting system violates IIA, multiple preference schedules have to be compared. Our IIA algorithm loops through all possible pairs of alternatives. For each alternative pair \(A\) and \(B\), the following test is done: First, find the societal preference between \(A\) and \(B\). Then, let \(j\) equal the number of voters who prefer \(A\) over \(B\) and \(k\) the number of voters who prefer \(B\) over \(A\). Then randomly generate 500 additional preference schedules in which \(j\) voters still prefer \(A\) over \(B\) and \(k\) voters prefer \(B\) over \(A\). The number 500 was chosen based on computational resources and run time.  If any of these new preference schedules produce a different societal preference between \(A\) and \(B\) than the original, an IIA violation occurs. This is repeated for every possible pair of alternatives.

This algorithm can miss violations, which is more likely when the number of voters is high. As the number of voters increases, there are more possible preference schedules. Thus, in the 500 new preference schedules generated, there may be no preference schedule that has a different societal order, even if one exists. Checking every single preference schedule is impractical since there are \((\frac{c!}{2})^v\) possible schedules such that each voter maintains their relative preference between a pair, where \(c\) is the number of alternatives and \(v\) is the number of voters. 

In Dougherty and Heckelman (2020), a different IIA algorithm was used. Our algorithm finds substantially more IIA violations, in particular with the IAC data. For Dowdall, Coombs, Copeland, and Baldwin our algorithm reports a greater percentage IIA violations. For Coombs and Copeland, the difference in the IAC data is around 20 percentage points. 

\subsubsection{Disagreement in IIA Conclusions under Copeland}

Dougherty and Heckelman (2020) conclude that Copeland cannot violate IIA if a majority of voters have identical preferences. However, consider an election with three voters and three alternatives with the following preference schedule where a strict majority have preference order \(A \succ B \succ C\):

 \begin{center}
\begin{tabular}{ c | c c c c c c }
  & 2 & 0 & 0 & 1 & 0 & 0 \\
   \hline
 1st & A & A & B & B & C & C \\
 2nd & B & C & A & C & A & B \\
 3rd & C & B & C & A & B & A \\
\end{tabular}
\end{center}

Under Copeland, the societal preference order is \(A \succ B \succ C\). Now, consider the pair of alternatives $A$ and $B$. Suppose that one voter who has preference \(A \succ B \succ C\) changes their preference to \(C \succ A \succ B\). Now the preference schedule is: 

\begin{center}
\begin{tabular}{ c | c c c c c c }
  & 1 & 0 & 0 & 1 & 1 & 0 \\
   \hline
 1st & A & A & B & B & C & C \\
 2nd & B & C & A & C & A & B \\
 3rd & C & B & C & A & B & A \\
\end{tabular}
\end{center}

The societal preference order is now \(A=B=C\). Even though all voters kept their relative preference between $A$ and $B$, the societal ranking between them changed. Copeland can violate IIA even if a majority have identical preferences.

Dougherty and Heckelman (2020) also state that Copeland violates IIA at a greater frequency for even number of voters \(N\) than for the adjacent odd values \(N - 1\) and \(N + 1\) as there are more ties for elections with an even number of voters. Our data refutes this claim, showing that IIA violations in 3-alternative elections are greater for elections with 3 voters (97.35\%) than with 4 voters (73.63\%).

\nocite{Xia}
\nocite{Xia_orig}

\end{document}